\documentstyle[galley,epsfig]{mn2e}

\title{Chemical analysis  of three barium  stars: HD~51959, HD~88035, HD~121447}
\title[Chemical analysis of three barium  stars]
{ Chemical analysis  of three barium  stars: HD~51959, HD~88035, HD~121447}
\author[Drisya Karinkuzhi et al. ]{Drisya Karinkuzhi$^{1}$, Aruna Goswami$^{1}$, Navin Sridhar$^{2}$, Thomas  Masseron$^{3}$,
\newauthor Meenakshi Purandardas$^{1}$  \\
    $^{1}$Indian Institute of Astrophysics, Koramangala, Bangalore 560034,
    India; drisya@iiap.res.in, aruna@iiap.res.in,meenakshi.p@iiap.res.in\\
$^{2}$ Indian Institute of Science Education and Research, 
 Bhopal Bhauri, Bhopal 462066, MP, India\\
$^{3}$  Institute of Astronomy, Madingley Road, Cambridge CB3 0HA, UK
}    
\begin{document}

\date{ Accepted ---;  Received ---;  in original 
form --- \large \bf }

\pagerange{\pageref{firstpage}--\pageref{lastpage}} \pubyear{2018}

%\maketitle

\label{firstpage}

\maketitle
\begin{abstract}
We present elemental abundance results from high resolution spectral 
analysis of three nitrogen-enhanced barium  stars. The analysis is 
based on spectra  obtained with  the FEROS attached to 1.52m telescope 
at ESO, Chile.  The spectral resolution  is R${\sim}$48000 and the 
spectral coverage spans from 3500-9000\AA\,.  For the objects HD51959 
and HD88035,  we present the first time abundance analyses results. 
Although a few studies are available in literature  on the object HD121447, 
the results are significantly different from each other. We have therefore
carried out  a detailed chemical composition study for this object 
 based on a high resolution spectrum with high S/N ratio, for a
 better understanding of  the origin of the abundance patterns 
observed in this star. Stellar atmospheric parameters,  the  
effective temperature, surface gravity, microturbulence and metallicity 
of the stars are determined from the LTE analysis using model atmospheres.  
The metallicity of HD51959 and HD88035 are found to be near-solar; 
they  exhibit enhanced abundances of neutron-capture elements. HD121447
is  found to be moderately metal-poor with [Fe/H]=$-$0.65. 
While carbon is near-solar in the other two objects, HD121447 
shows carbon enhancement at a level, [C/Fe]=0.82.   Neutron-capture
elements are highly enhanced  with [X/Fe]$>$2 (X: Ba, La, Pr, Nd, Sm)
in this object. The ${\alpha}$- and  iron-peak elements show abundances 
very similar to field giants with the same metallicity. From kinematic 
analysis all the  three  objects are found to be members of thin disk 
population  with  a high  probability of  0.99, 0.99 and 0.92 for HD51959,
HD88035 and HD121447 respectively.
 
\end{abstract}


\section{Introduction}
The chemically peculiar stars such as CH stars and barium stars that 
exhibit enhancement of slow neutron-capture elements in their 
surface chemical composition can provide observational constraints 
for models of neutron-capture nucleosynthesis that occur during 
the AGB phase of evolution of low and intermediate mass stars. Both 
the CH stars and barium stars are known to be binary systems 
(McClure 1983, 1984) with a now invisible white dwarf companion. 
The companions produced the neutron-capture elements during their 
AGB phase of evolution  and transferred these materials to the CH 
and barium stars through mass transfer mechanisms while evolving 
through the AGB phase. The mass transfer mechanisms are however 
not clearly understood. Although CH stars and barium stars show 
enhancement of slow neutron-capture elements and  very similar 
in this respect, there are a few properties that make them
distinct from each other. From kinematics, barium stars are 
known to be disk objects while CH stars belong to the halo 
of our Galaxy (Gomez et al. 1997, Mennessier et al. 1997). CH stars 
have high radial velocities and also they are metal-deficient 
(Hartwick and Cowley 1985). Also barium stars have longer
orbital period and greater eccentricities compared to the CH stars
(Vanture 1992a, Jorissen et al. 2016). Another distinguishing feature 
is the C/O ratio which is less than unity in case of barium stars
(Barbuy et al. 1992, Allen and Barbuy 2006, Drake and Pereira 2008, 
Pereira and Drake 2009) and greater than unity in case of CH 
stars (Vanture 1992b, Pereira and Drake 2009). Luck and Bond (1991) 
analyzed a few barium stars that show strong CH band but weak 
C$_{2}$ band and weak metallic lines. With the absence of strong
C$_{2}$ band in their spectra as they can not  be placed in the group
of CH stars, the authors categorized them as `metal-deficient-barium 
stars',  and referred them as population II analogs of classical 
barium stars. This thin line of difference raises questions 
concerning the evolutionary  connection, and the exact  
relationship between `metal-deficient barium stars' and the CH stars. 

The three chemically peculiar stars that are the subject of this 
present study are listed in the CH star catalogue of Bartkevicius 
(1996). However, as can be seen later our abundance analysis results
indicate these objects to be more likely   barium stars.  So far, 
no previous studies on chemical composition of the  stars  HD~51959 
and  HD~88035 exist in  literature.  
In section 2, we describe the source of the spectra of the 
programme stars and in section 3 we provide photometric temperatures
of the stars. In section 4, we discuss  abundance analysis and present
the abundance results.  In section 5 we have discussed briefly the 
kinematic analysis
of the stars. Some discussions and  conclusions are presented  in section 6.

\section{High resolution spectra of the programme stars}
High resolution spectra for the programme stars are obtained from the 
FEROS (Fiber-fed Extended Range Optical Spectrograph) connected to the 
1.52 m telescope at ESO, Chile. It covers the complete optical spectral 
range from 3500 - 9000 \AA\, in  a single exposure. The spectral resolution 
is R = 48000. FEROS is equipped with a 2K x 4K,  15${\mu}$m pixel CCD. 
An image slicer (Kaufer 1998) is used to achieve the high spectral 
resolution. The basic data for the  programme stars  are listed in  
Table 1. A few  sample spectra of  the programme stars are shown 
in figure 1.

\begin{figure}
\centering
\includegraphics[height=8cm,width=8cm]{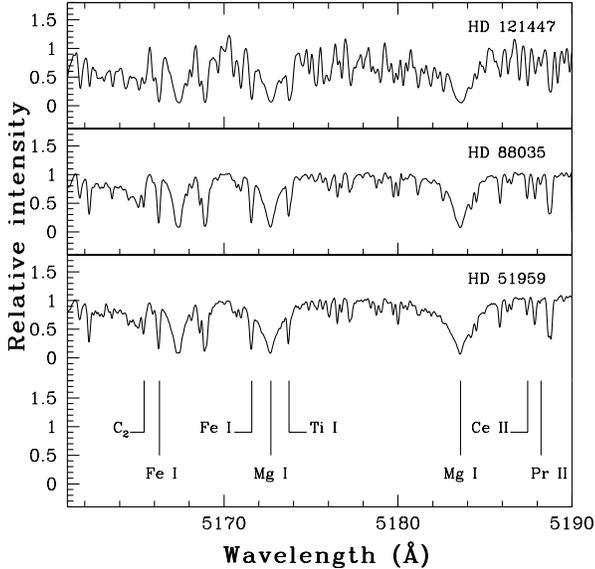}
\caption{ Sample spectra of the programme stars in  the wavelength 
region 5160 - 5190 \AA\,}
\end{figure}
%\label {Fig 1}

{\footnotesize 
\begin{table*}
{\bf Table 1: Basic data for the programme stars}\\
\begin{tabular}{ccccccccccc}
\hline
Star Name.   & RA(2000) & DEC(2000)& ${\pi}$(mas) & BC   &B& V&J&H&K& Date of obs \\
\hline
HD~51959   & 06 59 10.09&-07 06 31.95  &  8.05 [1.04] & $-$0.25  & 10.05 & 8.92 &  7.18 & 6.74 & 6.55&11-11-1999\\
 HD~88035  & 10 08 41.97&-20 18 49.64  &  2.78 [0.219]& $-$0.18  & 10.24 &  9.14 &  7.39 & 6.89 & 6.77&07-01-2000 \\
 HD~121447 &13 55 46.96  & -18 14 56.48& 2.93 [0.80] & $-$0.49  & 9.61 & 7.80 &   5.12 & 4.33 & 4.15&07-01-2000 \\
  
\hline
\end{tabular}

The numbers within square brackets indicate errors in ${\pi}$.
\end{table*}
}

{\footnotesize
\begin{table*}
{\bf Table 2: Temperatures from  photometry }\\
\begin{tabular}{llllllllllll}
\hline
Star Name  & $T_{eff}$  & $T_{eff}$ & $T_{eff}$ & $T_{eff}$ & $T_{eff}$& $T_{eff}$ &$T_{eff}$&$T_{eff}$&$T_{eff}$ &$T_{eff}$ & Spectroscopic \\
           &            & (-0.05)   & (-0.5)    & (-0.05)   & (-0.5)   & (-1.0)    &(-1.0)   & (-0.05)  & (-0.5)    &  (-1.0)    & estimates \\
           &  (J-K) K   & (J-H) K   & (J-H) K   & (V-K) K   & (V-K) K  &  (J-H) K  & (V-K) K & (B-V) K &(B-V) K  & (B-V) K & ~~~K \\
\hline
HD~51959   & 4564       &  4925     &  4968     &  4721     & 4720     & 4994      &  4714   & 4551    & 4467    & 4403  &5020  \\
HD~88035   & 4616       &  4673     &  4711     &  4726     & 4725     & 4732      &  4719   & 4604    & 4515    & 4446  & 5300 \\ 
HD~121447  & 3784       &  3775     &  3796     &  3905     & 3903     & 3806      &  3888   & 3616    & 3604    & 3607  &4500 \\

\hline
\end{tabular}

The numbers in the paranthesis indicate the metallicity values at which the 
temperatures are calculated.\\
\end{table*}
}
\section{Temperatures from photometric data}
The photometric temperatures of the programme stars are calculated 
as described in Mahanta et al. (2016) by using the calibration 
equations  for giants offered  by Alonso et al. (1999, 2001). 
The temperature estimates span a wide range. The photometric 
temperatures  are used as initial values in an iterative process 
to obtain the  spectroscopic temperature  estimates. While the 
spectroscopic temperature estimate for HD~51959 matches  closely 
with the temperature  estimate from (J-H) colour calibration, 
the spectroscopic temperatures  are much higher  than  those 
obtained from  (J-H) and (J-K) calibrations  in the case of other 
two objects.  We  note however  that  (J-H) and (J-K) calibrations 
give similar results in general as  in the case of HD~88035 and 
HD~121447 but  these two temperature estimates differ by about  
400 K in case of HD~51959. 
The reason for such a large difference  is difficult to understand if
the 2MASS J, H, K values for this object are considered to be  as 
accurate as for 
the other two stars. As strong  CH molecular absorption affects  the B band,    
we have not considered  the empirical T$_{eff}$ scale  for the 
B-V colour.  The JHK photometric temperature estimates were made 
to have a first hand temperature estimate for the stars. For our 
analysis we relied  on spectroscopic temperature estimates derived
using large numbers of clean Fe I and Fe II lines.
 The  estimated values along with the  spectroscopic temperatures are 
listed  in  Table 2.  

\section{Spectral analysis} 
We have derived the atmospheric parameters for the programme stars 
by measuring the equivalent widths  of clean unblended Fe I and 
Fe II lines. For this purpose  we have considered only those 
lines that have  excitation potential in the range   0.0 eV to 5.0 eV 
and  equivalent width between 20 m{\AA} to 160 m{\AA\,. For the  
object HD~51959  we could get  56 Fe I and  8 Fe II lines,  
for HD~88035 69 Fe I   and 7 Fe II lines, and,  for HD~121447 28 Fe I 
and 2 Fe II lines, that are usable for abundance 
calculation.  Due to severe line blending and distortion in the 
spectrum of HD~121447  we could not get sufficient  number of weak 
Fe lines within the above ranges of exitation potential and 
equivalent widths. We could notice a large broadening of spectral 
features throughout the spectrum of HD 121447. Hence, for this object  
we have also considered  a few strong unblended  lines for deriving 
the atmospheric parameters. We have used the latest version of 
MOOG (Sneden 1973) along with the Kurucz grid of model atmospheres 
with no convective overshooting 
(http: //cfaku5.cfa.harvard.edu/) for the calculations. We have assumed 
Local Thermodynamic Equilibrium (LTE) for the analysis.  Fe lines used 
for the analysis along with the measured equivalent widths  are 
presented in Table 3. The references for the adopted log\,gf values
are also listed in this table. The method of excitation equilibrium 
is used for deriving the effective  temperature T$_{eff}$, fixing at 
a  value that makes  the slope of the abundance versus excitation 
potential of Fe I lines   nearly  zero (Figure 2).  
The microturbulent velocity  at this temperature is fixed  by demanding 
that there be no dependence of the derived Fe I abundances on the 
equivalent width  of the corresponding lines (Figure 3). The surface 
gravity is fixed at a value which makes the abundance of Fe from 
the Fe I and Fe II lines equal. Derived  atmospheric parameters 
are presented in Table 4. 

We have also determined the surface gravity (log\,g) using the 
parallaxes (${\pi}$) from Van Leeuwen (2007) for HD~51959 and 
HD~121447. For the object HD~88035 we have adopted parallax value
from GAIA (http://archives.esac.esa.int/gaia). This method is 
precise when parallaxes have small measurement uncertainty. The 
following relation is used
   \begin{equation}
log {g \over g_{\odot}} = log {M \over M_{\odot}} + 4 log{T_{eff} \over T_{eff}{\odot}} + 0.4(M_{bol} - M_{bol}{\odot})  
\end{equation}
\vskip 0.2cm
\noindent
where
   \begin{equation}
 M_{bol} = V + 5 +5 log{\pi} +BC 
\end{equation}
The bolometric corrections are determined using the empirical calibration 
of Alonso et al. (1999) with its erratum Alonso et al. (2001) (Table 1).
We have adopted solar values
log\,$g_{\odot}$ = 4.44, T$_{eff}{\odot}$ = 5770 K and 
M$_{bol}{\odot}$ = 4.75 mag.
The masses of the programme stars  are derived from their locations 
in the Hertzsprung-Russell (HR) diagram using spectroscopic log T$_{eff}$ 
along the x-axis  
and photometric  log ($L/L_{\odot}$) along y-axis, where evolutionary tracks 
(Girardy et al. 2000) are plotted for  different masses (Figure 4). 
The estimated  
log ($L/L_{\odot}$) values of our stars are in good agreement  
with  estimates obtained for  Barium  giants by de Castro et al. (2016).

Errors in the mass estimates have maximum contribution coming from
the errors in parallaxes. Error in  parallaxes are 1.04, 0.219 and 0.8 
mas for HD 51959, HD 88035 and HD121447  respectively.  We have considered 
an error of 0.3 instead of 0.219  for HD~88035 since it is mentioned 
in the GAIA website  that errors below 0.3 are  very optimistic. The
errors in the log (L/L$_{\odot}$) estimates due to the errors in  
parallax values are   0.11, 0.09 and 0.24 for HD 51959, HD 88035 
and HD121447 respectively.

Since our objects are of near solar metallicities, we  have utilized  
the evolutionary tracks for the initial composition Z = 0.0198 
and Y = 0.273. The estimated masses 1.3 ${\pm}$ 0.1, 2.2 ${\pm}$ 0.1 and
2.0 ${\pm}$ 0.4 M$_{\odot}$ for HD~51959, HD~88035 and HD~121447
respectively are  presented in Table 5. The mass estimates are  well 
within the range of  model predictions  by Han et al. (1995)  who 
found that the masses of Ba stars  range between 1.0 and 3.0 M$_{\odot}$. 

The estimated log\,g from this method are   3.7 ${\pm}$ 0.09 (HD~51959),
 3.2 ${\pm}$ 0.03 (HD~88035) and  2.01 ${\pm}$ 0.05 (HD~121447).
The uncertainty in log g estimates are mostly due to uncertainty
in the parallax estimates that amounts to  about 12.9\% (HD~51959), 
7.9\% (HD~88035)  and 27.3\% (HD~121447). 
Although we 
have estimated log g values using  parallax method for a first check, 
we have used  spectroscopic estimates of log g for  our calculations. 

\begin{figure}
\centering
\includegraphics[height=8cm,width=8cm]{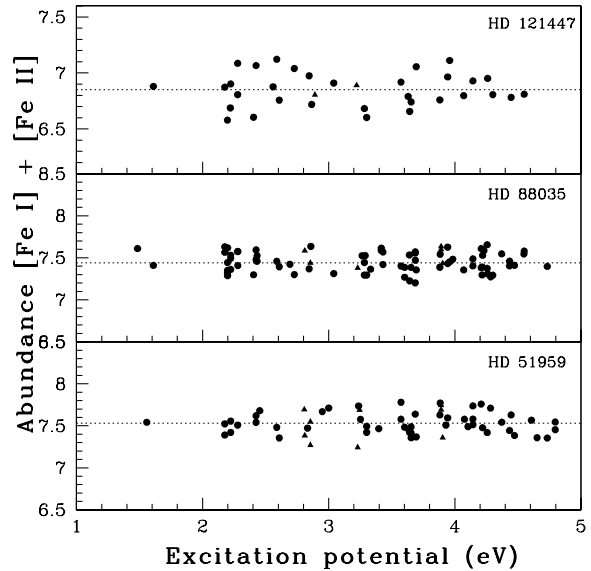}
\caption {The iron abundances of stars are shown for individual Fe I and
Fe II lines as a function of excitation potential. The solid circles 
indicate Fe I lines and solid triangles indicate Fe II lines. }

\end{figure}
%\label {Fig 2}

\begin{figure}
\centering
\includegraphics[height=8cm,width=8cm]{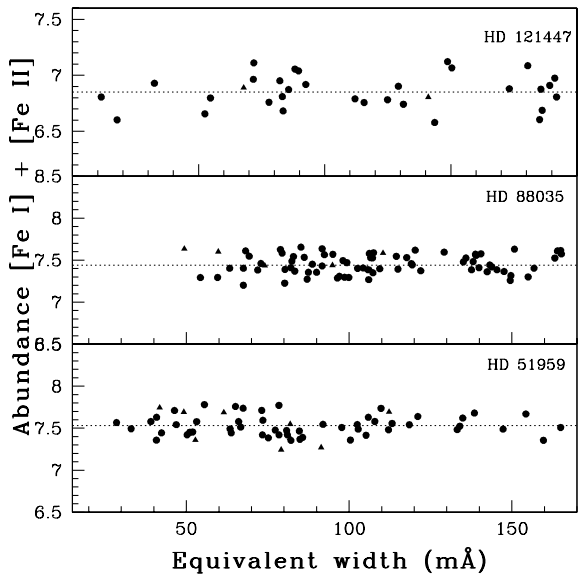}
\caption{ The iron abundances of stars are shown for individual Fe I and
Fe II lines as a function of equivalent width. The solid circles 
indicate Fe I lines and solid triangles indicate Fe II lines.}

\end{figure}
%\label {Fig 3}

{\footnotesize

\begin{table*}
{\bf  Table 3: Fe lines used for deriving atmospheric parameters}\\
\begin{tabular}{lccllllc}
\hline
Wavelength&   Element & E$_{low}$ & log gf   &  HD~51959 &HD~88035 &HD~121447 &Ref\\
~~~~{\AA} &   id      &  ev       &          &~~m{\AA}(dex)&~~m{\AA}(dex)&~~m{\AA}(dex)& \\
\hline
  4109.060 &  Fe I    &  3.290    & $-$1.56  & -        & 107(7.53) &  -        &1   \\
  4446.833 &          &  3.687    & $-$1.33  & -        &  99(7.47) &  -        &1 \\
  4447.129 &          &  2.198    & $-$2.59  &-         & 107(7.35) & -         &1  \\
  4447.720 &          &  2.220    & $-$1.34  &  -       &    -      & 236(6.69) &1\\
  4476.019 &          &  2.845    & $-$0.57  &-         & -         &  241(6.98)&2\\
  4485.971 &          &  3.654    & $-$2.35  & 41(7.41) &-          &           &1 \\
  4489.739 &          &  0.121    & $-$3.97  &-         & 157(7.40) &           &1  \\
  4566.514 &          &  3.301    & $-$2.25  & -        &  60(7.29) & 68(6.60)  &1  \\
  4614.192 &          &  3.301    & $-$2.62  & 50(7.49) & -         &           &1\\
  4619.287 &          &  3.602    & $-$1.12  & -        & 106(7.27) &  -        &1 \\
  
 \hline
\end{tabular}

The numbers in the paranthesis in columns 5 - 7 give the derived 
abundances from the respective line.\\
References : 1. F\"uhr et al. (1988) 2. Bridges $\&$ Kornblith (1974) \\
{\bf Note.}  This table is available in its entirety in online only. 
A portion is shown here for guidance regarding its form and content.\\
\end{table*}
}

\begin{table*}
{\bf  Table 4: Derived atmospheric parameters for the programme stars}\\
\begin{tabular}{lccccccc}
\hline
Star Name.   &  HJD    &V$_{r}$    & $T_{eff}$  &  log g  &  $\zeta $ &[Fe I/H] &[Fe II/H]    \\
             &         & km s$^{-1}$ &  K     &         &  km s$^{-1}$ &      &             \\
\hline
HD~51959& 2451493.837  &   38.5      & 5020 &3.65 &1.31 &0.03   &0.01\\
HD~88035& 2451550.715  &  $-$ 2.5    & 5300 &3.9 &1.88 &$-$0.05 &0.01\\
HD~121447& 2451550.805 &  $-$ 2.3    & 4500 &2.2 &3.02 &$-$0.65 &$-$0.65\\

\hline
\end{tabular}
\end{table*}

\begin{table*}
{\bf Table 5: Stellar masses}\\
\begin{tabular}{lllclcl}
\hline
Star Name & $M_{bol}$ & $ \Delta M_{bol}$ & log $(L/L_{\odot}$)& $\Delta {log (L/L_{\odot})}$& Mass ($M_{\odot}$) & $\Delta Mass(M_{\odot})$\\
\hline
HD 51959  & 3.19  &$\pm$0.26  &0.62 &$\pm$0.11& 1.3&$\pm$0.10\\
HD 88035  & 1.19  &$\pm$0.20  &1.43 &$\pm$0.09& 2.2&$\pm$0.10\\
HD 121447 & -0.35 &$\pm$0.52  &2.04 &$\pm$0.24& 2.0&$\pm$0.40\\
\hline
\end{tabular}
\end{table*}

\subsection{Elemental abundances }
 For most of the cases the  elemental abundances 
are derived by the standard method using the measured equivalent widths. 
For a few elements which are known to be affected by the hyperfine splitting, 
we have used the spectral  synthesis calculation to derive the abundances.  
 The elemental abundances along with the abundance ratios are presented 
in Tables  6 - 8. We have also calculated the [ls/Fe], [hs/Fe] and [hs/ls] 
values (Table 9), where ls represents light s-process elements Sr, Y and 
Zr and hs represents heavy s-process elements Ba, La, Ce,  Nd and Sm 
whenever available.  The lines used along with the abundances derived 
from the individual lines and the references for the log\,gf values adopted
for our calculations are presented in Table  10.   

\subsection{Uncertainties in metallicity and elemental abundances}
We have derived the uncertinities in elemental abundances by varying 
the stellar atmospheric  parameters T$_{eff}$, log g  and microturbulence 
in the model atmosphere. The  uncertainty due to temperature is determined
by varying the  temperatures by ${\pm}$ 100 K and recalculating 
the Fe abundance. Similarly by varying the log g value by ${\pm}$ 0.1 and 
microturbulent velocity by  ${\pm}$ 0.1 km s$^{-1}$, we have calculated 
the corresponding uncertainties in abundances due to these changes. 
The  total uncertainty is calculated using the standard equation  
of error calculations. We have assumed these values as the minimum 
error in the derived abundances. The  uncertainty in log ${\epsilon}$ 
is the standard error when they are derived from more than one line; 
when  abundances are  derived using  spectral synthesis calculations 
the uncertainty values are taken  as ${\pm}$ 0.2 dex which gives a clear
separation from adopted abundance value on either side of the observed
spectrum. The uncertainty in [X/H] can be considered  same as that 
for log ${\epsilon}$ since the adopted solar abundances are known 
to be precise  and  errors in the  solar abundances can be neglected. 
In the case of [X/Fe], the source of uncertainty  includes those 
from the measurement of [Fe/H] which adds up a minimum error 
of ${\sim}$ 0.10 dex. 

\subsection {Elemental abundance ratios: analysis  and  interpretation }
The estimated carbon in HD~121447 is found to be enhanced and near 
solar in the other two objects (Tables 6, 7, 8).   The carbon abundances  
are measured using spectrum  synthesis calculations of the weak  
C$_2$ band at 5165 \AA\, (figure 5).
Using this carbon abundance, the abundance of  nitrogen is  obtained 
from  spectrum synthesis calculations of the CN bands at 4215 \AA\, (figure 6)
 and 8005 \AA\,. 

 Nitrogen abundance derived from the CN red region 
(8005 \AA\,) is marginally higher by ${\sim}$ 0.15 dex in HD~88035 and 
HD~121447. For HD~51959, N abundance derived from 4215 \AA\, band is 
0.06 dex lower than that obtained from 8005 \AA\, region.
The abundances quoted in the tables 6, 7 and 8 are averages of these two 
values. Nitrogen is found to be enhanced in all 
the three objects. 
  For the linelists of C$_{2}$ and CN bands, we have 
consulted Brooke et al. (2013), Sneden et al. (2014) and Ram 
et al. (2014) and used the most updated log gf values for the 
C$_{2}$ lines in these regions.

 The abundances of oxygen in HD~51959 
and HD~88035 are estimated from spectrum synthesis calculation 
of the oxygen triplet  lines around   7774 \AA\, (figure 7). Although 
O I triplet 
is known to be affected by non-LTE conditions, the effect  
decreases in objects with higher gravities.  These effects  are also 
found to  disappear  for lines with  equivalent widths  below 
about 45 m\AA\,;  a discussion on these effects  for  K giants 
is available in Eriksson \& Toft (1979). 

Estimated  C/O  for HD~51959 and HD~88035 are   0.35 and
0.43 respectively are consistent with barium stars.
 The estimated  [ΣCNO/Fe]   for  HD~51959 and        
HD~88035  are found to be  0.20 and 0.39 respectively. 
From estimates of CNO abundances  Tomkin and Lambert (1979) have  
demonstrated in a sample of barium stars that carbon shows near solar 
values, oxygen is  mildly 
deficient ($\sim$ 0.1 dex), and,  nitrogen  is  mildly enhanced with 
$\sim$ 0.3 dex. However, Barbuy et al. (1992) have  found  carbon 
abundances in the range $-$0.25 $\leq$ [C/Fe] $\leq$ 0.3 for a
sample of barium stars; they  noted that the  less evolved 
Ba stars  show high N abundances (Barbuy et al. 1992, Allen and 
Barbuy 2006). 

Estimated   
$^{12}$C/$^{13}$C ratios obtained  using spectrum synthesis calculations
 of the   CN band at 8005 \AA\, (figure 8) are found to be
small with  values  10.1 and 7.3 for  HD~88035 and HD~121447 respectively.
 The  $^{12}$C/$^{13}$C ratios are derived using the set of $^{12}$CN 
lines at 8003.553, 8003.910 \AA\,  and  $^{13}$CN features at 8004.554, 
8004.728, 8004.781 \AA\, which are considered to be more reliable 
for this calculation. 

\begin{figure}
\centering
\includegraphics[height=8cm,width=8cm]{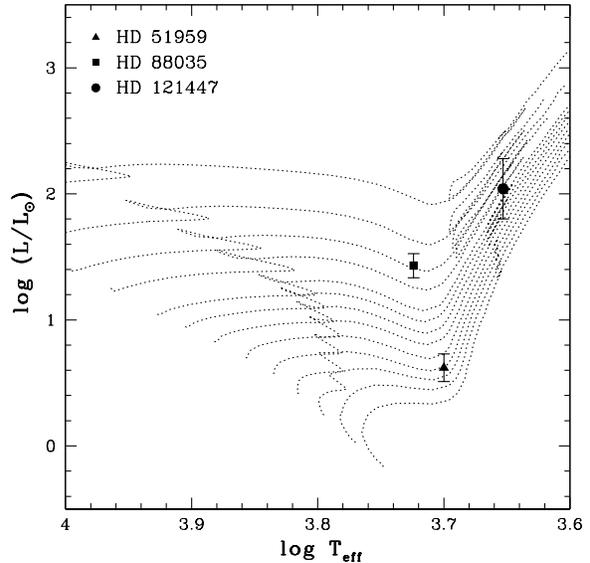}
\caption{The locations of HD 51959, HD~88035  and HD 121447 are shown
 the Hertzsprung-Russell diagram. The evolutionary tracks from Girardi et al.
 (2000) are shown for masses  1.0, 1.1, 1.2, 1.3, 
1.4, 1.5, 1.6, 1.7, 1.8, 2.0, 2.2, 2.5 and 3.0 M$_{\odot}$ from bottom to top. }
\end{figure}
%\label{Fig 4}

\begin{figure}
\centering
\includegraphics[height=8cm,width=8cm]{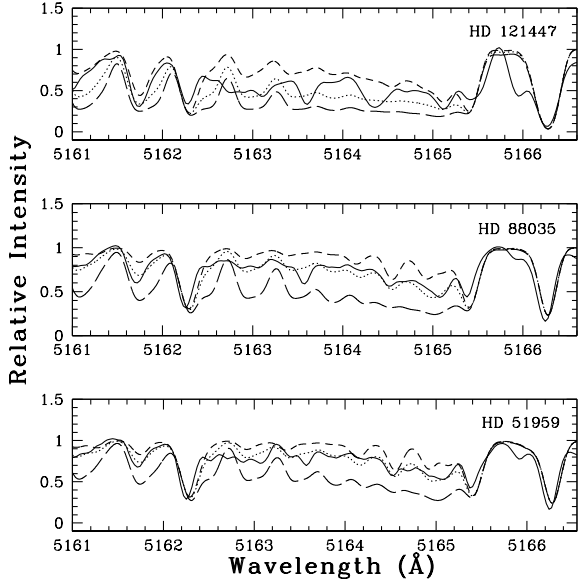}
\caption{
The spectral synthesis of $C_2$ band around  5165  \AA\,. In all the 
panels the dotted lines indicate the synthesized spectra and the solid
lines indicate the observed line profiles. Two alternative synthetic 
spectra for [C/Fe] = +0.3 (long-dashed line) and [C/Fe] = $-$0.3 
(short-dashed line) are shown to demonstrate the sensitivity of 
the line strength to the abundances.  }
\end{figure}
%\label {Fig 5}

\begin{figure}
\centering
\includegraphics[height=8cm,width=8cm]{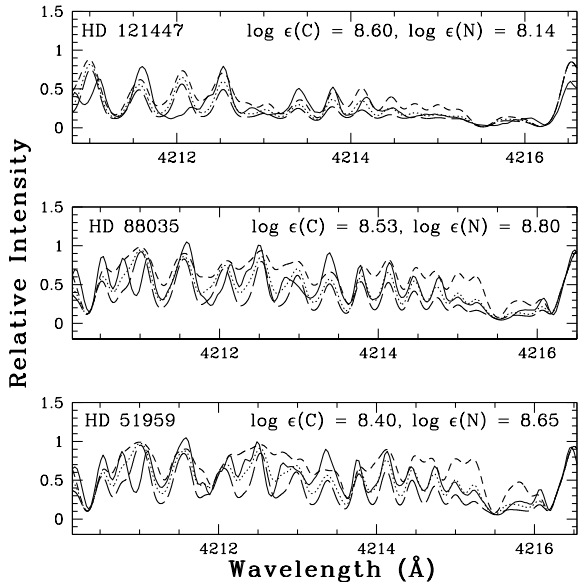}
\caption{ Spectral synthesis fits of CN band around  4215 \AA\,. 
The best fit obtained  with  a carbon abundance of 8.7 dex and 
$^{12}$C/$^{13}C$ ${\sim}$ 10   returns a nitrogen abundance 
of 7.1 dex  (dotted lines). The solid line corresponds to the 
observed spectrum. Two alternative plots with long-dash and 
short-dash are shown with [N/Fe] = $\pm$0.3 from the adopted value. 
}
\end{figure}
%\label{Fig 6}

\begin{figure}
\centering
\includegraphics[height=8cm,width=8cm]{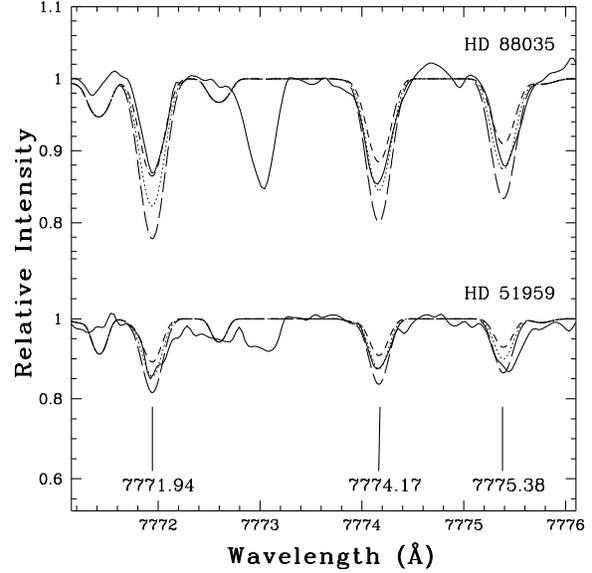}
\caption{ The spectral synthesis plots of OI triplet lines obtained with the
adopted O abundances (dotted curve). The observed spectrum is shown 
by a solid curve. Two alternative plots with long-dash and short-dash 
are shown with [O/Fe] = $\pm$0.3 from the adopted value.
}
\end{figure}
%\label {Fig 7}

\begin{figure}
\centering
\includegraphics[height=8cm,width=8cm]{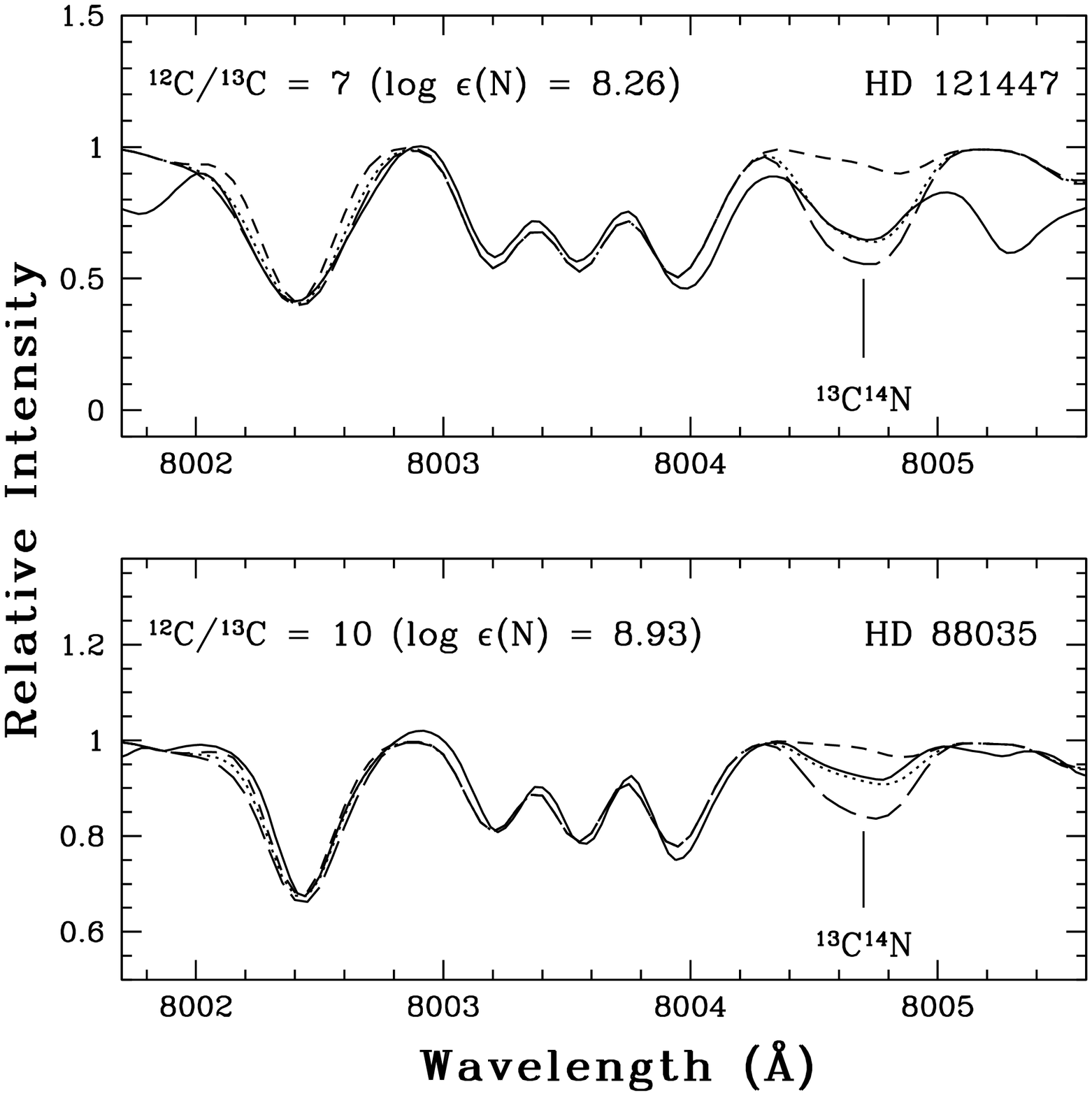}
\caption{ The spectral synthesis fits of the CN features around  8005  \AA\,
obtained with  the adopted N abundance and  $^{12}$C/$^{13}$C values
(dotted curve). The observed spectrum is shown by  a solid curve.
 Two alternative fits with  the solar value of $^{12}$C/$^{13}$C ${\sim}$ 90
(short-dashed line) and 5 (long-dashed line) are  shown to illustrate
the sensitivity of the line strengths to the isotopic carbon  abundance
ratios.
}
\end{figure}
%\label {Fig 8}

 The near solar Na abundances (Tables 6 \& 7)  
 derived using the Na I lines at 5682.65 and 5688.22 \AA\
 for HD~51959 and HD~88035  are as normally seen in 
field giants.
The sample of barium stars studied by de Castro 
et al. (2016)  show  [Na/Fe] ratios  in the range 0.3 to 0.6. 

A single Al I line at 6783.68 \AA\,  gives [Al/Fe] ${\sim}$
 0.49 dex for HD~51959, a value  not so uncommon to barium stars; 
de Castro et al.
(2016) found [Al/Fe] ${\sim}$ 0.43 for HD~43389, which is a  metal-poor
star  with [Fe/H] ${\sim}$ $-$0.52.

Except for Mg that shows mild underabundance,  the abundance ratios 
of ${\alpha}$- elements in HD~51959 and HD~88035
are very similar to those normally seen in giants and barium stars.
Mg is also found to be underabundant 
in case of 
 HD~49641 and HD~58368, two strong barium stars analyzed
by Mahanta et al. (2016). 
  The lowest [Mg/Fe] found in the sample
of de Castro et al. (2016) is [Mg/Fe] ${\sim}$ 0.13 for HD~142751
([Fe/H] = $-$ 0.1). Our estimate of [Mg/Fe]  for HD~121447 is
near solar with   value of  0.04.

Silicon and calcium  show near solar values for HD~51959 and HD~88035
(Tables 6 \& 7), these  values in  HD~121447  are marginally higher than that 
normally seen in barium stars. 

The abundance  of Ti in case of  HD~51959 is   near solar (Table 6), and
 marginally higher  
 in HD~88035 and HD~121447. These values are consistent with  
 those  of de Castro's sample.

 Abundance of Scandium is  derived for   HD~51959 and HD~88035
 using spectrum synthesis calculation  of Sc II line at 6245.63 {\AA}  
considering hyperfine structure from Prochaska $\&$ McWilliam (2000). 
 [Sc/Fe]  is near solar in HD~51959 and mildly 
enhanced in HD~88035 (Tables 6, 7).  The abundance of Sc could not
be measured in HD~121447. 

Abundance of V is  estimated using spectrum synthesis calculation 
of V I  line at 5727.028 {\AA\,} taking into account the hyperfine
components from Kurucz data base (Table 6). 
Yang et al. (2016) finds a similar value 
of [V/Fe] = 0.2 for HD~81797. Their  sample  shows [Sc/Fe]
in the range $-$0.07 to 0.17.   Vanadium  is  mildly enhanced in HD~88035 
and HD~121447 with [V/Fe] =  0.42 (Tables 6 and 7)
Abundances of Sc and V are not reported in de Castro et al.  (2016).

 Abundance ratios of iron peak elements in HD~51959 and HD~88035 
exhibit similar values as seen in normal giants and  barium stars. 
Cr shows near solar values in all the three objects (Tables 6, 7, 8). 

The abundance of manganese is derived using spectrum synthesis calculation 
 of 6013.51 {\AA}  line taking into account the 
hyperfine structures from Prochaska $\&$ McWilliam (2000). Considering 
the uncertainty limits,  Mn is found to be near solar in HD~51959
 and mildly underabundant in HD~88035 (Tables 6 \& 7).
This ratio is  +0.51 for 
HD~121447 which is marginally above the values normally seen 
in barium stars. The abundance of Mn in Yang et al. (2016)
ranges from $-$0.27 (HD~58121) to 0.3 (HD~43232) in their sample.

While Co shows a mild enhancement in HD~51959 and HD~88035,
the  abundances  of Ni are near solar   (Tables 6 \& 7).
A mild 
enhancement is noticed in HD~121447 with a value of 0.37.

Abundance of Zn derived using a single line at 4810.528 {\AA\,}
gives [Zn/Fe] ${\sim}$ 0.21 for  HD~88035.
Abundances of Mn, Co and Zn in  de Castro et al. (2016) and Co and Zn in 
Yang et al. (2016) are  not reported. 

Abundance of  Sr is measured for  HD~51959 and HD~88035 
using both the equivalent width measurements as well as spectrum synthesis
calculation of the Sr I line at 4607.7 {\AA} and found to be enhanced.
 Abundances of Sr are not reported for their sample in
de Castro et al. (2016) and Yang et al. (2016).

Abundance of Y  derived from the Y II lines indicates enhancement 
 in all the three stars. 
 Spectrum synthesis calculation fits  for  Y II line 
at 5289.81 {\AA\,} is shown in figure 9.
For HD~121447, the Y abundance  is derived using two Y II lines at 
5119.11 and 5544.61 {\AA}.

Zirconium abundance derived from Zr I lines in HD~88035 and HD~121447
and  Zr II lines at 4414.54 
and 4613.97 {\AA\,}  for HD~51959 show large enhancement.
Enhancement of Y and Zr
abundances are noticed in the sample of Yang et al. (2016) and also in the
sample of four barium stars studied by Mahanta et al. (2016).
The sample of de Castro et al. (2016) shows a wide range in abundance ratios
from near solar to [X/Fe] $>$ 1.0 for both Y and Zr.

\begin{figure}
\centering
\includegraphics[height=8cm,width=8cm]{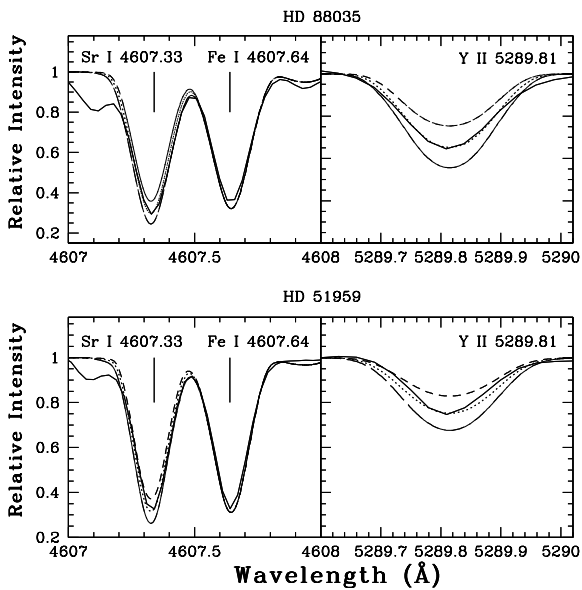}
\caption{Spectral synthesis of Sr I line at 4607.33 {\AA} and Y II 
line at  5289.81 {\AA} are shown for HD 51959 (lower panel) and HD 88035 
(upper panel). The dotted lines indicate the synthesized spectra and 
the solid lines indicate the observed line profiles. Two alternative 
synthetic spectra for [X/Fe] = +0.3 (long-dashed line) and 
[X/Fe] = $-$0.3 (short-dashed line) from the observed value are shown
to demonstrate the sensitivity  of the line strength to the abundances. }
\end{figure}
%\label{Fig 9}

  We have used Ba II lines at 4934, 5853, 6141 and 6496 {\AA} 
whenever available to derive the  barium sbunadnce.
Spectrum synthesis calculation  of Ba II line at 5853 {\AA} 
(figure 10, left panels) 
 considering the hyperfine splitting contributions from Mc William (1998)
is also performed to estimate barium abundance. 
Reported [Ba/Fe] values  in Tables 6-8 are 
those obtained from  spectrum synthesis calculations  of 
Ba II feature at 5853.66 {\AA}.  While
de Castro et al. (2016) have not reported abundances of  Ba for their sample,
[Ba/Fe] ranges from 0.17 (HD~11658) to 1.13 (HD~49641) in the sample of
Yang et al. (2016).
 
Abundance of lanthanum is derived  from the spectrum synthesis calculation
 of La II line at 4921.77 {\AA\,} 
(figure 10, right panels) 
considering hyperfine components from Jonsell et al. (2006). The 
derived values of La  indicate large enhancements (Tables 6, 7, 8).
Estimates of [La/Fe] range from 0.26 (HD~11658) to 1.38 (HD~49641) 
in the sample of Yang et al. (2016); these estimates have a wider range
for the sample of de Castro et al. (2016) from 0 (BD$-$01302) to 
2.7 (HD~24035). 

\begin{figure}
\centering
\includegraphics[height=8cm,width=8cm]{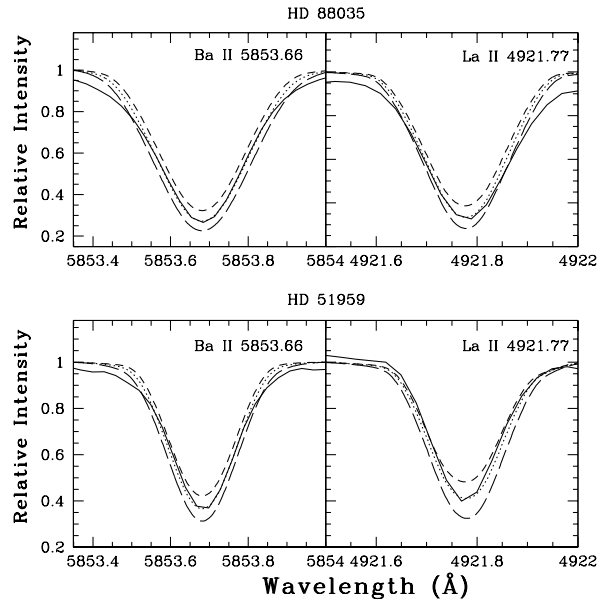}
\caption{Spectral synthesis of Ba II line at 5853.66 {\AA\,} and 
La II line at 4921.77 {\AA\,} are shown for HD 51959 (lower panel) 
and HD 88035 (upper panel). The dotted lines indicate the synthesized 
spectra and the solid lines indicate the observed line profiles. 
Two alternative synthetic spectra for [X/Fe] = +0.3 (long-dashed line) 
and [X/Fe] = $-$0.3 (short-dashed line) from the observed value 
are shown to demonstrate the sensitivity  of the line strength 
to the abundances. }
\end{figure}
%\label{Fig 10}

 Ce, Pr, Nd and Sm abundances are also found to be enhanced in all
the three stars (Tables 6, 7, 8). 
 In the sample of
de Castro et al.  the estimates of  [Ce/Fe] 
 range from $-$0.08 (HD~212484) to
1.87 (HD~107541) and [Nd/Fe] 
  from 0.0 (HD~51315) to 1.83 (HD~107541).

A spectrum synthesis calculation
 of Eu II line at 6645.130 {\AA\,}
 by considering the hyperfine components from Worley et al.
(2013) shows enhancement of Eu  in HD~121447. 
Since this line is found to be slightly  distorted,
our estimate may be regarded as  an upper limit. 
 For HD~88035, the Eu abundance 
derived from the spectrum synthesis calculation   of  Eu II line 
at 6437.64 {\AA} 
shows an  enhancement with [Eu/Fe] ${\sim}$ 0.96.  

 Abundance of dysprosium  is derived  using Dy II lines
at 4103.310 and 4923.167 {\AA\,} for HD~51959.
For HD~88035 and  HD~121447, Dy abundance was derived from a single line 
at 4923.17 {\AA}.

Estimated  [hs/ls] $>$ 0 for HD~88035 and HD~121447 (where hs refers 
to the second peak s-process elements and ls refers to the first 
peak s-process 
elements) is an indication that the neutron exposures  experienced 
in their AGB progenitor companions are sufficiently strong for the
production  of the more abundant `hs' elements than the `ls' elements.
 
{\footnotesize
\begin{table*}
{\bf Table 6 : Elemental abundances in HD~51959}\\
\begin{tabular}{|l|c|c|c|c|c|}
\hline
 &       &                         &                &        &        \\
 &    Z  &   solar $log{\epsilon}^a$ & $log{\epsilon}$& [X/H]  & [X/Fe] \\
 &       &                         &      dex            &        &      \\
\hline
C {\sc i}  &   6      &   8.43  &  8.40$\pm$0.20(syn)  &   -0.03  &  -0.06 \\
N {\sc i}  &   7      &   7.83  &  8.68$\pm$0.20(syn)  &    0.85  &  0.82 \\
O {\sc i}  &   8      &   8.69  &  8.85$\pm$0.20(syn)  &    0.16  &  0.13 \\
Na {\sc i}  &  11     &   6.24   &  6.29$\pm$0.17(6)  &   0.05  &  0.02 \\
Mg {\sc i}  &  12     &   7.60   &  7.21$\pm$0.16(2)  &   -0.39  &  -0.42  \\
Al {\sc i}  &  13     &   6.45   &6.97$\pm$0.20(1) & 0.52  &  0.49  \\
Si {\sc i}  &  14     &   7.51   &  7.67$\pm$0.21(4)  &    0.16  &  0.13\\
Ca {\sc i}  &  20     &   6.34   &  6.11$\pm$0.16(5)  &   -0.23  &  -0.26\\
Sc {\sc ii}* &  21     &   3.15   &   3.30$\pm$.20(1,syn)  &  0.15   &  0.14\\
Ti {\sc i}  &  22     &   4.95   &  4.89$\pm$0.15(13)  &   -0.06  &  -0.09\\
Ti {\sc ii} &  22     &   4.95   &  4.93$\pm$0.20(1)  &   -0.02 & -0.03      \\
V {\sc i}*   &  23     &   3.93   &  4.10$\pm$0.20(1,syn)   &    0.17 & 0.14\\
Cr {\sc i}  &  24     &   5.64   &  5.54$\pm$0.18(10)   &  -0.10  &  -0.13\\
Mn {\sc i}*  &  25     &   5.43   &  5.20$\pm$0.20(1,syn)   &  -0.23  & -0.26\\
Fe {\sc i}  &  26     &   7.50   &  7.53$\pm$0.12(56)  &  0.03& -\\
Fe {\sc ii}  &  26     &   7.50   &  7.51$\pm$0.20(8)  &  0.01& -\\
Co {\sc i}  &  27     &   4.99   &  5.20$\pm$0.13(6)  &    0.21      & 0.24\\
Ni {\sc i}  &  28     &   6.22   &  6.19$\pm$0.12(14)  &   -0.03  &  -0.07\\
Sr {\sc i}*  &  38     &   2.87   &  4.18$\pm$0.20(1,syn)  &   1.31  & 1.28\\
Y {\sc ii}*  &  39     &   2.21   &  3.20$\pm$0.18(5)  &   0.99  &  0.98\\
Zr {\sc ii} &  40     &   2.58   &  3.86 $\pm$0.24(2)   &   1.28  & 1.25\\
Ba {\sc ii}* &  56     &   2.18   &  3.00$\pm$0.20(1,syn)   &   0.82  & 0.81\\
La {\sc ii}* &  57     &   1.10   &  2.05$\pm$0.20(1,syn)  &   0.95  & 0.94\\
Ce {\sc ii} &  58     &   1.58   &  2.61$\pm$0.21(12)  &   1.03  & 1.02\\
Pr {\sc ii} &  59     &   0.72   &  1.91$\pm$0.17(3) &   1.19  & 1.18\\
Nd {\sc ii} &  60     &   1.42   &  2.59$\pm$0.20(9)  &   1.17  & 1.16\\
Sm {\sc ii} &  62     &   0.96   &  1.99$\pm$0.13(5)  &   1.03  & 1.02\\
Dy {\sc ii}  &  66     &   1.10  & 2.71$\pm$0.22(2) &   1.61   &  1.60\\
\hline 
\end{tabular}

$^{a}$ Asplund et al. (2009) \\
$^*$ abundances are derived using spectral synthesis of respective lines.\\
\end{table*}
}

{\footnotesize
\begin{table*}
{\bf Table 7  : Elemental abundances in HD~88035}\\
\begin{tabular}{|l|c|c|c|c|c|}
\hline
 &       &                         &                &        &      \\
 &    Z  &   solar $log{\epsilon}^a$ & $log{\epsilon}$& [X/H]  & [X/Fe]\\
 &       &                         &       dex           &        &       \\
\hline
C {\sc i}  &   6      &   8.43  &  8.53$\pm$0.20(syn)  &   0.10  &  0.1\\
N {\sc i}  &   7      &   7.83  &  8.87$\pm$0.20(syn)  &   1.04  &  1.09 \\
O {\sc i}  &   8      &   8.69  &  8.90$\pm$0.20(syn)  &   0.21  &  0.26 \\
Na {\sc i}  &  11     &   6.24   &  6.36$\pm$0.13(4)  &    0.12  &  0.17\\
Mg {\sc i}  &  12     &   7.60   &  7.23$\pm$0.03(2)  &   -0.37  &  -0.32  \\
Si {\sc i}  &  14     &   7.51   &  7.69$\pm$0.13(2)  &   0.18  &  0.13\\
Ca {\sc i}  &  20     &   6.34   &  6.24$\pm$0.16(8)  &   -0.10  &  -0.05\\
Sc {\sc ii}* &  21     &   3.15   &  3.50$\pm$0.20(1,syn)  &   0.35  &  0.34\\
Ti {\sc i}  &  22     &   4.95   &  5.06$\pm$0.16(13)  &   0.11  & 0.16\\
Ti {\sc ii} &  22     &   4.95   &  5.22$\pm$0.18(8)  &    0.27  &  0.26\\
V {\sc i}*   &  23     &   3.93   &  4.30$\pm$0.20(1,syn)  &   0.37  & 0.42\\
Cr {\sc i}  &  24     &   5.64   &  5.56$\pm$0.19(6)  &   -0.08  &  -0.03\\
Mn {\sc i}*  &  25     &   5.43   &  4.95$\pm$0.20(1,syn)  &   -0.48  & -0.43\\
Fe {\sc i}  &  26     &   7.50   &  7.45$\pm$0.11(69)  &  -0.05& -\\
Fe {\sc ii} &  26     &   7.50   &  7.51$\pm$0.10(7)  &  0.01& -\\
Co {\sc i}  &  27     &   4.99   &  5.36$\pm$0.20(5) &    0.37  & 0.42\\
Ni {\sc i}  &  28     &   6.22   &  6.24$\pm$0.18(20)  &   0.02   &  0.07\\
Zn {\sc i}  &  30     &   4.56   &  4.30$\pm$0.20(1)  &   0.26   &  0.21  \\
Sr {\sc i}*  &  38     &   2.87   &  4.27$\pm$0.20(1,syn)  &   1.40   & 1.45\\
Y {\sc ii}*  &  39     &   2.21   &  3.42$\pm$0.14(5)  &   1.21   &  1.20\\
Zr {\sc i}  &  40     &   2.58   &  3.83$\pm$0.11(4)  &   1.25  &1.30 \\
Ba {\sc ii}* &  56     &   2.18   &  3.70$\pm$0.20(1,syn)  &   1.52  & 1.51\\
La {\sc ii}* &  57     &   1.10   &  2.80$\pm$0.20(1,syn)  &   1.70  & 1.69\\
Ce {\sc ii} &  58     &   1.58   &  3.36$\pm$0.19(10)  &   1.78  & 1.76\\
Pr {\sc ii} &  59     &   0.72   &  2.70$\pm$0.05(3)  &   1.98  & 1.97\\
Nd {\sc ii} &  60     &   1.42   &  3.17$\pm$0.16(11)  &   1.75  & 1.74\\
Sm {\sc ii} &  62     &   0.96   &  2.90$\pm$0.13(5)  &   1.94  & 1.93\\
Eu {\sc ii}* &  63     &   0.52   &  1.49$\pm$0.19(1,syn)  &   0.97   & 0.96\\
Dy {\sc ii}  &  66     &   1.10  & 2.56$\pm$0.20(1) &   1.46   &  1.45\\
\hline 
\end{tabular}

$^{a}$ Asplund et al. (2009) \\
$^*$ abundances are derived using spectral synthesis of respective lines.\\
\end{table*}
}

{\footnotesize
\begin{table*}
{\bf Table 8  : Elemental abundances in HD~121447}\\
\begin{tabular}{|l|c|c|c|c|c|}
\hline
 &       &                         &                &        &        \\
 &    Z  &   solar $log{\epsilon}^a$ & $log{\epsilon}$& [X/H]  & [X/Fe] \\
 &       &                         &        dex          &        &        \\
\hline
C {\sc i}  &   6      &   8.43  &  8.60$\pm$0.20(syn)  &   0.17  &  0.82 \\
N {\sc i}  &   7      &   7.83  &  8.26$\pm$0.20(syn)  &   0.43  &  1.08 \\
Mg {\sc i}  &  12     &   7.60   &  6.99$\pm$0.20(1)  &   -0.61  &  0.04 \\
Si {\sc i}  &  14     &   7.51   &  7.73$\pm$0.20(1)  &   -0.04  &  0.61\\
Ca {\sc i}  &  20     &   6.34   &  6.30$\pm$0.18(4)  &   -0.04  &  0.61\\
Ti {\sc i} &  22     &   4.95   &  4.59$\pm$0.02(3)  &    -0.36  &  0.29\\
V {\sc i}*   &  23     &   3.93   &  4.30$\pm$0.20(1,syn)  &   0.37  & 0.42\\
Cr {\sc i}  &  24     &   5.64   &  5.15$\pm$0.05(2)  &   -0.49  &  0.16\\
Mn {\sc i}*  &  25     &   5.43   &  5.29$\pm$0.20(1,syn)  &   -0.14  & 0.51\\
Fe {\sc i}  &  26     &   7.50  &  6.85$\pm$0.15(28)  &  -0.65& -\\
Fe {\sc ii} &  26     &   7.50   &  6.85$\pm$0.06(2)  &  -0.65& -\\
%Co {\sc i}  &  27     &   4.99   &  5.61$\pm$0.13(3) &    0.62  & 1.27\\
Ni {\sc i}  &  28     &   6.22   &  5.94$\pm$0.18(6)  &   -0.28   &  0.37\\
Y {\sc ii}*  &  39     &   2.21   &  3.17$\pm$0.26(2)  &   0.96   &  1.61\\
Zr {\sc i}  &  40     &   2.58   &  3.62$\pm$0.12(2)  &   1.04  &1.69\\
Ba {\sc ii}* &  56     &   2.18   &  4.20$\pm$0.20(1,syn)  &   2.02  & 2.67\\
La {\sc ii}* &  57     &   1.10   &  2.40$\pm$0.20(1,syn)  &   1.67  & 2.32\\
Ce {\sc ii} &  58     &   1.58   &  2.60$\pm$0.16(4)  &   1.02  & +1.67\\
Pr {\sc ii} &  59     &   0.72   &  2.62$\pm$0.06(5)  &   1.90  & 2.55\\
Nd {\sc ii} &  60     &   1.42   &  2.89$\pm$0.16(5)  &   1.47  & 2.12\\
Sm {\sc ii} &  62     &   0.96   &  2.48$\pm$0.15(3)  &   1.52  & 2.17\\
Eu {\sc ii}* &  63     &   0.52   &  1.30$\pm$0.20(1,syn)  &   0.78  & 1.43\\
Dy {\sc ii}  &  66     &   1.10  & 2.96$\pm$0.20(1) &   1.86   &  2.51\\
\hline 
\end{tabular}

$^{a}$ Asplund et al. (2009) \\
$^*$ abundances are derived using spectral synthesis of respective lines.\\
\end{table*}
}

\begin{table*}
{\bf Table 9: Observed values for [Fe/H], [ls/Fe], [hs/Fe]  and [hs/ls]}\\
\begin{tabular}{lcccc}
\hline
Star Name & [Fe/H]  & [ls/Fe] & [hs/Fe] & [hs/ls]   \\
\hline
HD 51959& 0.03& 1.18& 1.00& -0.18\\
HD 88035& -0.05& 1.29& 1.67& 0.14\\
HD 121447 & -0.65&1.63&2.39&0.76 \\
\hline
\end{tabular}
\end{table*}

{\footnotesize
\begin{table*}
{\bf  Table 10: lines used for deriving elemental abundances}\\
\begin{tabular}{lccllllc}
\hline
Wavelength&    Element    &    E$_{low}$ &   log gf&  HD~51959&HD~88035&HD~121447&Ref\\
~~~~~ {\AA}   &      id        &  ev &         &   ~~m{\AA}(dex)       & ~~m{\AA}(dex)        &   ~~m{\AA}(dex)&     \\       
\hline
  5682.650 &  Na I&   2.100  & -0.70  &  140(6.52) &   136(6.46)     & - &1\\  
  5688.220 &      &   2.100  & -0.40  &  139(6.21) &   146(6.27) 	  &-&1\\
  6160.747 &      &   2.105  & -1.26  &   85(6.31) &    -         	  &-&1\\
  8194.824 &      &   2.105  &  0.49  &  258(6.12) &    -         	  &-&1\\
  5889.950 &       &  0.000  &  0.10  &  545(5.55) &   626(5.87)	  &-&2\\
  5895.920 &      &   0.000  & -0.20  &  601(5.95) &   471(5.87) 	  &-&2\\
  4702.990 &  Mg I&   4.350  & -0.67  &  229(7.37) &   214(7.26) 	  &-&3\\
  5528.400 &      &   4.350  & -0.49  &  224(7.22) &   227(7.20)  &265.9(6.98)&3\\ 
  6783.680 &  Al I&   4.021  & -1.44  &  45(6.98)  &    -	  &-&4\\
  4782.990 &  Si I&   4.954  & -2.47  &     -      &    38(7.85)	  &-&1\\
\hline
\end{tabular}

The numbers in the paranthesis in columns 5 - 7 give the derived 
abundances from the respective line.\\
References:   1. Kurucz \& Peytremann (1975), 2. Wiese et al. (1966), 
3. Lincke \& Ziegenbein (1971), 4. Kurucz (1975)\\
{\bf Note.}  This table is available in its entirety in online only. 
A portion is shown here for guidance regarding its form and content.\\
\end{table*}
}

  \begin{table*}
{\bf  Table 11: Abundance Uncertainities}\\
\begin{tabular}{llllll}
\hline
Star & Standard  & $\delta T_{eff}$&$ \delta$log g &$ \delta \xi$ & Total\\
     &    error  & $\pm$ 100K     & $\pm$ 0.1 dex  & $\pm$ 0.3 km s$^{-1}$ &  Uncertainity \\  
\hline
HD 51959 & 0.12 & 0.06 & 0.04 & 0.03&0.14\\
HD 88035 & 0.11& 0.03 & 0.02 & 0.03&0.12\\
HD 121447&0.15& 0.04  &0.02&0.01&0.16\\

\hline
\end{tabular}
\end{table*}
{\footnotesize 
\begin{table*}
{\bf  Table 12: Atmospheric parameters of  HD 121447 }\\
\begin{tabular}{lccccc}
\hline
Star Name.   & $T_{eff}$  &  log g  &  $\xi $ &[Fe /H] &Ref \\
             &   K        &         &  km s$^{-1}$ &      &             \\
\hline
HD~121447&4500&2.2&3.02&-0.65&1\\
         &4000&1.0&2.0& -0.25&2\\
         &3900&1.0&2.0&-0.50& 3\\
         &4200&0.8&2.5&0.05&4\\
          
\hline
\end{tabular}

1. Our work, 2. Abia et al. (1998), 3. Merle et al. (2016), 4. Smith (1984)\\
\end{table*}
}

\section {Kinematic analysis} 

We have calculated the space velocity for the stars  with respect to 
the Sun using the  method of  Johnson \& Soderblom (1987). The space 
velocity with respect to the Local Standard of Rest (LSR) is  given by  \\

 \begin{center}
  $(U, V, W)_{LSR} =(U,V,W)+(U, V, W)_{\odot}$ km/s.
  \end{center}

 where, $(U, V, W)_{\odot} =(11.1, 12.2, 7.3)$ km/s (Schonrich et al., 2010) 
is the solar motion with respect to LSR and  
   \begin{center}
  $\left[ \begin{array}{c}  U \\  V \\  W \end{array} \right] = B.\left[ \begin{array}{c} V_{r}  \\ k.\mu_{\alpha}/\pi  \\ k.\mu_{\delta}/\pi \end{array} \right]$ 
  \end{center}
  where, $B=T.A$, T is the transformation matrix connecting the Galactic 
coordinate system and equatorial coordinate system, k = 4.74057 km s$^{-1}$
equivalent of 1 AU in one year and $\mu_{\alpha}$ and $\mu_{\delta}$ are
proper motions in RA and Dec.

The estimates of the components of spatial velocity, $U_{LSR}$, 
$V_{LSR}$ and $W_{LSR}$ measured along axes pointing towards 
the Galactic center, the direction of Galactic rotation and the North 
Galactic Pole respectively are with respect to the Local Standard 
of Rest (LSR).  We have used the  radial velocity estimate and the 
corresponding error estimate as  measured by us. The estimates for 
proper motion are taken from SIMBAD. Distances are measured taking
parallax values from SIMBAD and GAIA whenever available.

\noindent 
The total spatial velocity of a star is given by,\\
  $V_{spa}^{2}=U_{LSR}^{2}+V_{LSR}^{2}+W_{LSR}^{2}$

The  estimated components of spatial velocity and the total spatial 
velocity are presented in Table 13. According to Chen et al. (2004) 
V$_{spa}$ $\le$ 85 km s$^{-1}$ for a thin disk star. The total 
spatial velocities for all the three stars are found to be below 
this value.

We have also  calculated the probability for the stars to be a member 
of thin disk, the thick disk or the halo population following  the 
procedures of  Reddy et al.(2006), Bensby et al. (2003, 2004) and 
Mishenina  et al. (2004). 

The estimated metallicities and low spatial velocities indicate the 
stars to be thin disk objects. The probability estimates for them
being members of the thin disk population are 0.99, 0.99 and 0.92 
for HD~51959, HD~88035 and HD~121447 respectively (Table 13).

\begin{table*}
{\bf Table 13: Spatial velocity and probability estimates for the program stars.}
  \begin{tabular}{|l|l|l|l|l|l|l|l|}
  \hline\hline
   Star name & $U_{LSR}$ (km/s) & $V_{LSR}$ (km/s) & $W_{LSR}$ (km/s) & $V_{spa}$ (km/s) & $p_{thin}$ & $p_{thick}$ & $p_{halo}$\\
   \hline
 HD 51959 & $-25.17\pm0.98$ & $-4.12\pm1.2$ & $-9.58\pm2.19$ & $27.25\pm1.86$ & 0.99 & 0.009 & 0.00019\\
 HD 88035 & $-36.00\pm4.06$ & $13.02\pm0.84$ & $-12.24\pm2.02$ & $40.19\pm4.52$ & 0.988 & 0.01079 & 0.0002\\
 HD 121447 & $-45.19\pm15.05$ & $-17.72\pm8.54$ & $35.61\pm8.19$ & $60.21\pm18.60$ & 0.922 & 0.0755 & 0.0023\\
   \hline
  \end{tabular}
   \end{table*}

{\footnotesize 
\begin{table*}
{\bf  Table 14: Abundance ratios of carbon and neutron-capture elements in
 HD~121447 }\\
\begin{tabular}{lcccccccccccc}
\hline
Star Name.   & [C/Fe] & [N/Fe]  &  [Sr/Fe]  &  [Y/Fe]  &[Zr/Fe] & [Ba/Fe]& [La/Fe]& [Ce/Fe] & [Nd/Fe]& [Sm/Fe]& [Eu/Fe] & Ref \\
\hline
HD 121447& 0.82& 0.96 & -& 1.61&1.64&2.67&2.33&1.67&2.12&2.17&1.43&1\\
         &- & -& 1.2&1.5&1.15&1.3&2.0&1.5&1.3&-&0.95&2\\
         &0.55& - &0.97&0.94&1.86&2.33&1.74&1.27&-&-&-&3\\
         &0.22& 0.52 & 1.22&0.76&0.66&0.57&0.70&0.73&0.65&-&0.39&4\\
\hline
\end{tabular}

1. Our work, 2. Abia et al. (1998), 3. Merle et al. (2016), 4. Smith (1984)\\
\end{table*}
}

\section{Discussions and conclusions}
 Chemical analysis of three stars from the Bartkevicius (1996) catalogue 
of CH stars clearly shows spectral properties that are characteristics 
of barium stars. In particular, with [Ba/Fe] ${\sim}$ 0.81, the object 
HD~51959  seems to be a mild barium star, as assigned by 
Udry et al. (1998a, 1998b) and Jorissen et al. (1998),
HD~88035 a potential member of strong barium stars, and HD~121447 a 
metal-poor barium star.

CH stars are known to be high velocity objects with 
V$_{r}$r $\ge$ $\pm$100 km s$^{-1}$ . None of the three objects is found 
to be a high velocity object (Table 4). From kinematic analysis all the 
three stars are found to be members of thin disk population with high 
probability. From radial velocity variations Jorissen et al. (1998, 2016) 
have measured a period of 185.7 days and eccentricity 0.01 for the object
HD~121447. 
These values are in consistence with the majority of barium and CH stars. 
They have suggested this object to be an ellipsoidal variable. The 
orbital information for HD~88035 is not available in literature. HD~51959 
is believed to be a long period ($\sim$9488 days) object with an 
eccentricity of 0.58 (Jorissen et al. in preparation). This is not 
surprising as barium stars are known to have longer orbital periods 
and greater eccentricities than their population II analogs, the CH stars.

All the three objects in our sample show enhancement in Nitrogen 
with [N/Fe] in the range 0.6 to 1.0 (figure 11). An insightful 
discussion on C, N, O abundances of barium stars can be found in 
Smith (1984). For the object HD~121447, that is common in his sample, 
the best values for C and O were found to be logN(C) = 8.7 
and logN(O) = 8.8. Using these abundances of C and O he had derived 
Nitrogen abundance to be logN(N) = 8.4 for HD~121447. Our C and N 
abundance values  are slightly lower  with log N(C) = 8.6 
and log N(N) = 8.14 for this object. We note however, that our 
stellar parameters for this object differ considerably with those 
of Smith (1984) (Table 12).  As discussed by many authors 
(Smith (1984), Luck and Lambert (1985), Barbuy et al. (1992), Smiljanic 
et al. (2006), Merle et al. (2016)), higher N abundances with low C 
abundances observed in some Ba stars indicate CN processing. Increase 
in [N/C] ratio as the star ascends the giant branch is attributed to 
the mixing process like first dredge up (FDU). Smiljanic et al. (2006) 
explains about a more complex kind of rotational mixing which can cause 
the increase in [N/C] ratio. From the HR diagram, our objects are in the 
first ascend of the giant branch, which clearly supports increase in 
N abundance over Carbon's. 
Light elements abundance in Ba stars are expected to scale with the 
metallicity values.  From figure 12, it is clear that our 
results are consistent with those for  other objects with similar 
metallicity values. 

The  heavy elements abundance ratios observed in the three stars
match closely with majority of barium stars in literature. This is clearly 
evident  from a  comparison of the estimates of the abundance ratios with
those of de Castro et al. (2016) and Yang et al. (2016) (figure 13).

The objects HD~88035 and HD~121447 are peculiar with very high N abundance 
along with large enhancement of heavy-s process elements. This could 
possibly be due to the presence high mass companions to these objects 
where the neutron densities are expected to be high. The large N 
enhancement in these objects could be due to HBB operating in the higher 
mass AGB companions of these objects. Moreover, higher enhancement of 
r-process elements Sm, Eu and Dy in these objects could also indicate 
the presence of higher neutron densities in their AGB companions 
(Cowan and Rose 1977). But the derived low masses for these  objects 
which represents a lower limit to the companion mass does not support 
the presence of higher mass companion. 

\begin{figure}
\centering
\includegraphics[height=8cm,width=8cm]{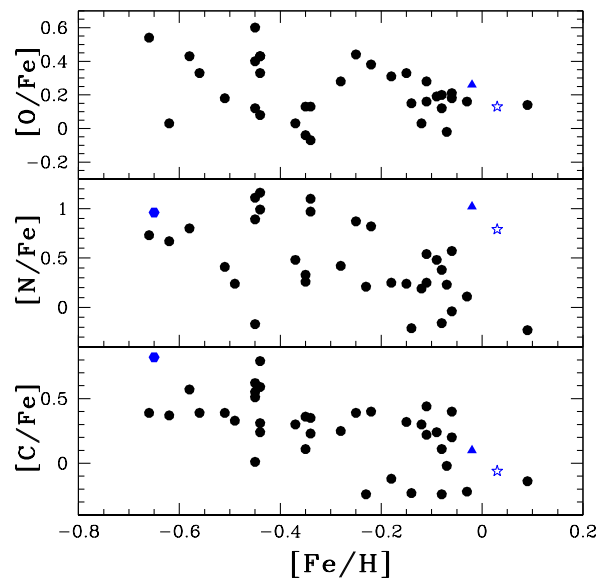}
\caption{CNO abundance ratios observed in HD 51959 (star symbol), HD 88035 (solid tringle) and HD 121447(solid hexagon) with respect to metallicity [Fe/H].
 Solid circles represent Ba stars from literature (Smith (1984), Barbuy (1992),
 Allen and Barbuy (2006), Merle et al. (2016)).}
\end{figure}
%\label {Fig 11}

\begin{figure}
\centering
\includegraphics[height=8cm,width=8cm]{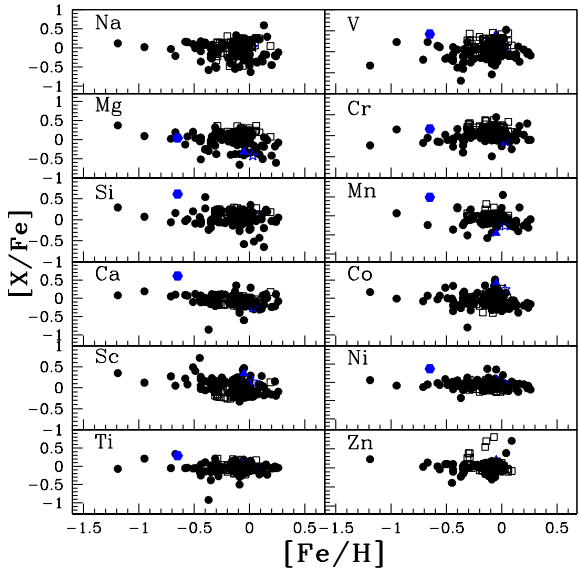}
\caption{Abundance ratios of light elements observed in 
HD 51959 (star symbol), HD 88035 (solid tringle) and HD 121447 (solid hexagon) 
with respect to metallicity [Fe/H]. Solid circles represent Ba stars from 
literature (Allen and Barbuy (2006), Liang et al. (2003),
Smiljanic et al. (2007), Zacs (1994), de Castro et al. (2016) and 
Mahanta et al. (2016)). Open squares represent 
normal giants from literature (Luck and Heiter 2007, 
Mishenina et al. 2006). }
\end{figure}
%\label {Fig 12}

\begin{figure}
\centering
\includegraphics[height=8cm,width=8cm]{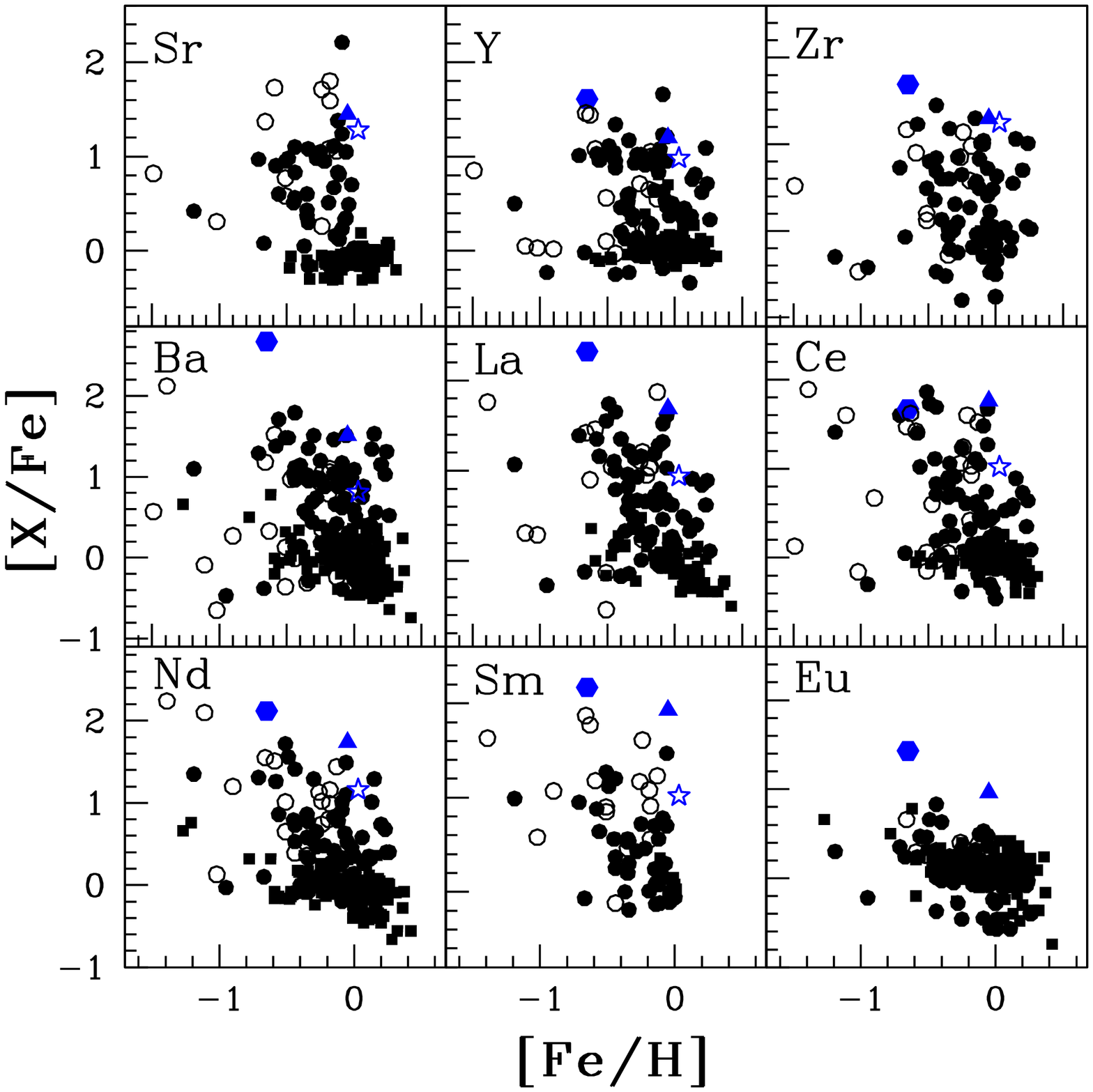}
\caption{Abundance ratios of heavy elements observed in HD 51959 
(star symbol), HD 88035 (solid tringle) and HD 121447 (solid hexagon) 
with respect to metallicity [Fe/H]. Solid circles represent Ba stars from 
literature (Allen and Barbuy (2006), Liang et al. (2003), Smiljanic et al. 
(2007), Zacs (1994), de Castro et al. (2016) and Mahanta et al. (2016)). 
Solid squares represent normal giants from literature (Van der Swaelman 
et al. (2016), Tautvaisiene et al. (2000) and Luck and Heiter (2007)). 
Open circles indicate CH stars from Karinkuzhi \& Goswami (2014, 2015). }
\end{figure}
%\label{Fig 13}
\vskip 0.3cm
\noindent
{\it Acknowledgement}\\
This work made use of the
SIMBAD astronomical database, operated at CDS, Strasbourg, France, and the
NASA ADS, USA.  M. P. is a JRF in the
DST project SB/S2/HEP-010/2013; funding from this project
is gratefully acknowledged. N Sridhar, a student of integrated BS-MS 
course of IISER, Bhopal, gratefully acknowledges the summer internship 
at IIA, under IIA's Visiting Student Program - 2015. \\

\end{document}